\documentclass[conference,final]{IEEEtran}
\IEEEoverridecommandlockouts

\usepackage{amsmath,amssymb,amsfonts}
\usepackage{algorithmic}
\usepackage{graphicx}
\usepackage{textcomp}

\usepackage{hyperref}

\usepackage[svgnames,x11names,table]{xcolor}

\usepackage{enumitem}

\usepackage{xspace}

\usepackage{subcaption}

\usepackage{tikz}
\usetikzlibrary{arrows.meta}
\usetikzlibrary{shapes.geometric}

\usepackage[nameinlink,noabbrev]{cleveref}

\usepackage[style=ieee,url=false,isbn=false,doi=false,eprint=false]{biblatex}

\hypersetup{
  colorlinks=true,
  citecolor=DarkGreen,
  linkcolor=NavyBlue,
  urlcolor=DarkRed,
  filecolor=green,
  anchorcolor=black,
  unicode=true,
  pdfencoding=unicode,
  plainpages=false,
  breaklinks=true,
  pdfauthor={Renato Cordeiro Ferreira},
  pdftitle={Reusability in MLOps: Leveraging Ports and Adapters to Build a Microservices Architecture for the Maritime Domain},
  pdfkeywords={MLOps, Ports and Adapters, Reusability, Software Architecture, Machine Learning-Enabled Systems, Maritime Domain, Experience Report},
}

\emergencystretch=\maxdimen
\definecolor{drawio-blue}{HTML}{6c8ebf} %
\definecolor{drawio-gray}{HTML}{666666} %
\definecolor{drawio-green}{HTML}{82b366} %
\definecolor{drawio-orange}{HTML}{fa6800} %
\definecolor{drawio-pink}{HTML}{99004D} %
\definecolor{drawio-purple}{HTML}{9673a6} %
\definecolor{drawio-red}{HTML}{b85450} %
\definecolor{drawio-white}{HTML}{f9f7ed} %
\definecolor{drawio-black}{HTML}{36393d} %
\definecolor{drawio-yellow}{HTML}{d6b656} %

\definecolor{drawio-violet}{HTML}{6a00ff} %
\definecolor{drawio-magenta}{HTML}{dd0073} %
\definecolor{drawio-moss}{HTML}{008a00} %

\newlength{\FramedXMargin}
\newlength{\FramedYMargin}
\newlength{\FramedSkipAbove}
\newlength{\FramedSkipBelow}

\newsavebox{\FramedBox}

\newenvironment{Framed}{%
  \begin{lrbox}{\FramedBox}
  \begin{minipage}{\dimexpr\linewidth-2\fboxrule-2\fboxsep} 
  \vspace{\FramedYMargin}%
  \centering%
  \begin{minipage}{\dimexpr\linewidth-2\FramedXMargin} 
}{%
  \end{minipage}%
  \vspace{\FramedYMargin}
  \end{minipage}%
  \end{lrbox}%
  \vspace{\FramedSkipAbove}
  \noindent\fbox{\usebox{\FramedBox}}
  \vspace{\FramedSkipBelow}
}

\newcounter{RQ}

\newcounter{SubRQ}
\renewcommand{\theSubRQ}{\arabic{SubRQ}}

\crefformat{RQ}{#2RQ#3}
\Crefformat{RQ}{#2RQ#3}
\crefformat{SubRQ}{#2sRQ#1#3}
\Crefformat{SubRQ}{#2sRQ#1#3}

\crefrangeformat{SubRQ}{#3sRQ#1#4 to #5sRQ#2#6}
\Crefrangeformat{SubRQ}{#3sRQ#1#4 to #5sRQ#2#6}

\newcommand{\TikzSkewedSquare}[1][black]{%
  \tikz[baseline] {\draw[transform shape, black, fill={#1}] (0,0) rectangle (1.5ex,1.5ex);}%
}
\newcommand{\TikzSkewedCircle}[1][black]{%
  \tikz[baseline] {\draw[yshift=0.75ex, anchor=south, transform shape, black, fill={#1}] (0,0) circle (0.75ex);}%
}
\newcommand{\TikzSkewedHexagon}[1][black]{%
  \tikz[baseline] {\node[yshift=0.75ex, regular polygon, regular polygon sides=6, shape border rotate=0, draw=black, fill={#1}, scale=0.8] (hex) {}}%
}

\newcommand{\LegendColoredSquare}[3]{%
  \textcolor{#1}{#3}~\TikzSkewedSquare[#2]%
}
\newcommand{\LegendColoredCircle}[3]{%
  \textcolor{#1}{#3}~\TikzSkewedCircle[#2]%
}
\newcommand{\LegendColoredHexagon}[3]{%
  \textcolor{#1}{#3}~\TikzSkewedHexagon[#2]%
}

\newcommand{\LegendColoredComponent}[2]{%
  \LegendColoredSquare{drawio-#1}{drawio-#1}{#2}%
}
\newcommand{\LegendBWComponent}[1]{%
  \LegendColoredSquare{drawio-black}{drawio-white}{#1}%
}

\newcommand{\LegendColoredLabel}[2]{%
  \LegendColoredCircle{drawio-#1}{drawio-#1}{#2}%
}
\newcommand{\LegendColoredLayer}[2]{%
  \LegendColoredHexagon{drawio-#1}{drawio-#1}{#2}%
}

\newcommand{\LegendService}[2]{%
  \textcolor{drawio-violet}{{#2~(#1)}}%
}
\newcommand{\LegendPipeline}[2]{%
  \textcolor{drawio-moss}{{#2~(#1)}}%
}
\newcommand{\LegendDataStore}[2]{%
  \textcolor{drawio-magenta}{{#2~(#1)}}%
}

\newcommand{\OG}{\textsc{Ocean Guard}\xspace}

\newcommand{\Subsystem}[1]{\textbf{#1}\xspace}

\newcommand{\Service}[1]{\texttt{#1}\xspace}

\newcommand{\Adapter}[1]{\texttt{#1}\xspace}

\newcommand{\Pattern}[1]{\textsc{#1}\xspace}

\makeatletter
\newcommand{\linebreakand}{%
  \end{@IEEEauthorhalign}
  \hfill\mbox{}\par
  \mbox{}\hfill\begin{@IEEEauthorhalign}
}
\makeatother

\begin{document}


\title{
Reusability in MLOps:\break%
Leveraging Ports and Adapters\break%
to Build a Microservices Architecture\break%
for the Maritime Domain%
\thanks{
This study was financed in part by the Coordenação de Aperfeiçoamento
de Pessoal de Nível Superior -- Brasil (CAPES) -- Finance Code 001.
It also received support from the MARIT-D project, co-funded from the Internal
Security Fund -- Police programme under grant agreement no. 101114216.
}
}


\author{

\IEEEauthorblockN{Renato Cordeiro Ferreira}
\IEEEauthorblockA{\textit{Jheronimus Academy of Data Science (JADS)} \\
\textit{Tilburg University (TiU)}\\
's-Hertogenbosch, The Netherlands \\
0000-0001-7296-7091}

\and

\IEEEauthorblockN{Aditya Dhinavahi}
\IEEEauthorblockA{\textit{Jheronimus Academy of Data Science (JADS)} \\
\textit{Technical University of Eindhoven (TU/e)}\\
's-Hertogenbosch, The Netherlands \\
0009-0004-5203-551X}

\linebreakand

\IEEEauthorblockN{Rowanne Trapmann}
\IEEEauthorblockA{\textit{Jheronimus Academy of Data Science (JADS)} \\
\textit{Technical University of Eindhoven (TU/e)}\\
's-Hertogenbosch, The Netherlands \\
0000-0001-5746-4154}

\and

\IEEEauthorblockN{Willem-Jan van den Heuvel}
\IEEEauthorblockA{\textit{Jheronimus Academy of Data Science (JADS)} \\
\textit{Tilburg University (TiU)}\\
's-Hertogenbosch, The Netherlands \\
0000-0003-2929-413X}

}

\maketitle


\begin{abstract}
ML-Enabled Systems (MLES) are inherently complex since they require multiple
components to achieve their business goal. This experience report showcases the
software architecture reusability techniques applied while building \OG, an MLES
for anomaly detection in the maritime domain. In particular, it highlights the
challenges and lessons learned to reuse the \Pattern{Ports and Adapters} pattern
to support building multiple microservices from a single codebase. This
experience report hopes to inspire software engineers, machine learning
engineers, and data scientists to apply the \Pattern{Hexagonal Architecture}
pattern to build their MLES.
\end{abstract}




\section{Introduction}
\label{sec:introduction}

Maritime transport is important for the functioning of global trade and
commerce. Operations in the maritime domain are dynamic and complex for multiple
reasons, including vulnerable weather and sea conditions, movement across
international boundaries, and compliance with varying regulatory frameworks%
~\cite{Charamis:2025:TrendsShippingIndustry}. 

The maritime industry faces the challenge of illicit trade, which is a threat to
peace and security~\cite{Bueger:2015:MaritimeSecurity,
Kavallieratos:2020:SecurityRequirementsShips}.
Among existing products that identify potential illegal activity at sea,
the \OG tool is an extensible Machine Learning--Enabled System (MLES) whose goal
is to analyze and detect anomalies across multiple types of data%
~\cite{Ferreira:2025:OceanGuard:MLOpsWithMicroservices}.


This experience report describes the software architecture reusability
techniques used to develop the \OG tool. It builds upon our first description
of the system~\cite{Ferreira:2025:OceanGuard:MLOpsWithMicroservices},
detailing how our development team applied the \Pattern{Ports and Adapters}
pattern to implement multiple microservices of the MLES from a single
monorepo codebase. 




\section{Fundamentals}
\label{sec:fundamentals}

This section introduces the key concepts and techniques used throughout the
paper.

\subsection{Machine Learning--Enabled Systems (MLES)}
\label{subsec:mles}

MLES are software systems that include components based on Machine Learning (ML)
as part of their business logic~\cite{Huyen:DesigningMLSystems:2022,
Lwakatare:2020:ChallengesSolutionsMLES}. \OG is an MLES because it
uses ML to implement anomaly detection techniques over maritime data.

MLES are complex systems because they rely on multiple components
to achieve their task~\cite{Huyen:DesigningMLSystems:2022}.
The usage of different component types and their most common roles has
been summarized in different reference architectures available in the
literature~\cite{Kreuzberger:2023:MLOpsOverview,
Kumara:2025:MLOpsReferenceArchitecture}.

The \OG tool follows the reference architecture proposed by
\citeauthor{Ferreira:2025:PhD:MetricsOrientedArchitecturalModel}%
~\cite{Ferreira:2025:PhD:MetricsOrientedArchitecturalModel},
as detailed in~\cref{sec:system_architecture}.

\subsection{Hexagonal Architecture}
\label{subsec:hexagonal_architecture}

The \Pattern{Hexagonal Architecture} pattern, also known as the
\Pattern{Ports and Adapters} pattern, was first described by
\citeauthor{Cockburn:2005:HexagonalArchitecture}%
~\cite{Cockburn:2005:HexagonalArchitecture}.
Its goal is to decouple the application's core business logic from external
dependencies~\cite{Martin:CleanArchitecture:2017}, such as storage or presentation.

This pattern is commonly described for implementing microservices%
~\cite{Newman:BuildingMicroservices:2021,Richardson:MicroservicesPatterns:2018}.
The code of an application that follows the \Pattern{Hexagonal Architecture}
pattern usually can be divided in three layers:
\begin{itemize}
    \item \textbf{core},
           which implements the business logic, usually by following Domain-Driven
           Design patterns~\cite{Evans:DomainDrivenDesign:2003};
    \item \textbf{ports},
           which defines contracts for the interaction between
           the core and external dependencies; and
    \item \textbf{adapters},
           which implement the contracts defined by the ports,
           communicating with the external dependencies.
\end{itemize}
Moreover, such an application requires an entry point that can connect these
components via \Pattern{Dependency Injection}~\cite{Martin:CleanArchitecture:2017}.

The \OG tool has microservices implemented using the \Pattern{Hexagonal Architecture}
pattern, as detailed in \cref{sec:application_architecture}.
\section{Related Literature}
\label{sec:related_literature}

Few works in the literature focus on the use of microservices for MLOps.
Searching with the keywords ``Maritime Microservices with MLOps'' on
Google Scholar does not provide any meaningful related results.

A search for the keywords ``Microservices with MLOps'' and
``Maritime Microservices'' results in few related results.
\citeauthor{Liu:2022:BlueNaviMaritimeMicroservices}%
~\cite{Liu:2022:BlueNaviMaritimeMicroservices}
proposes the use of microservices to efficiently serve maritime data,
but includes no anomaly detection on its components.
\citeauthor{Roh:2023:MicroservicesMLOps}%
~\cite{Roh:2023:MicroservicesMLOps}
proposes the use of microservices architecture to stably deploy and maintain
systems running ML models but does not address the whole MLOps lifecycle
nor focus on the maritime domain.

\section{\OG}
\label{sec:ocean_guard}

\OG is an MLES for anomaly detection in the maritime domain. The design of the
system follows a microservices architectural style to allow multiple teams to
develop in parallel~\cite{Ferreira:2025:OceanGuard:MLOpsWithMicroservices}.

\begin{figure*}[p]
  \centering
  \includegraphics[width=0.83\linewidth]{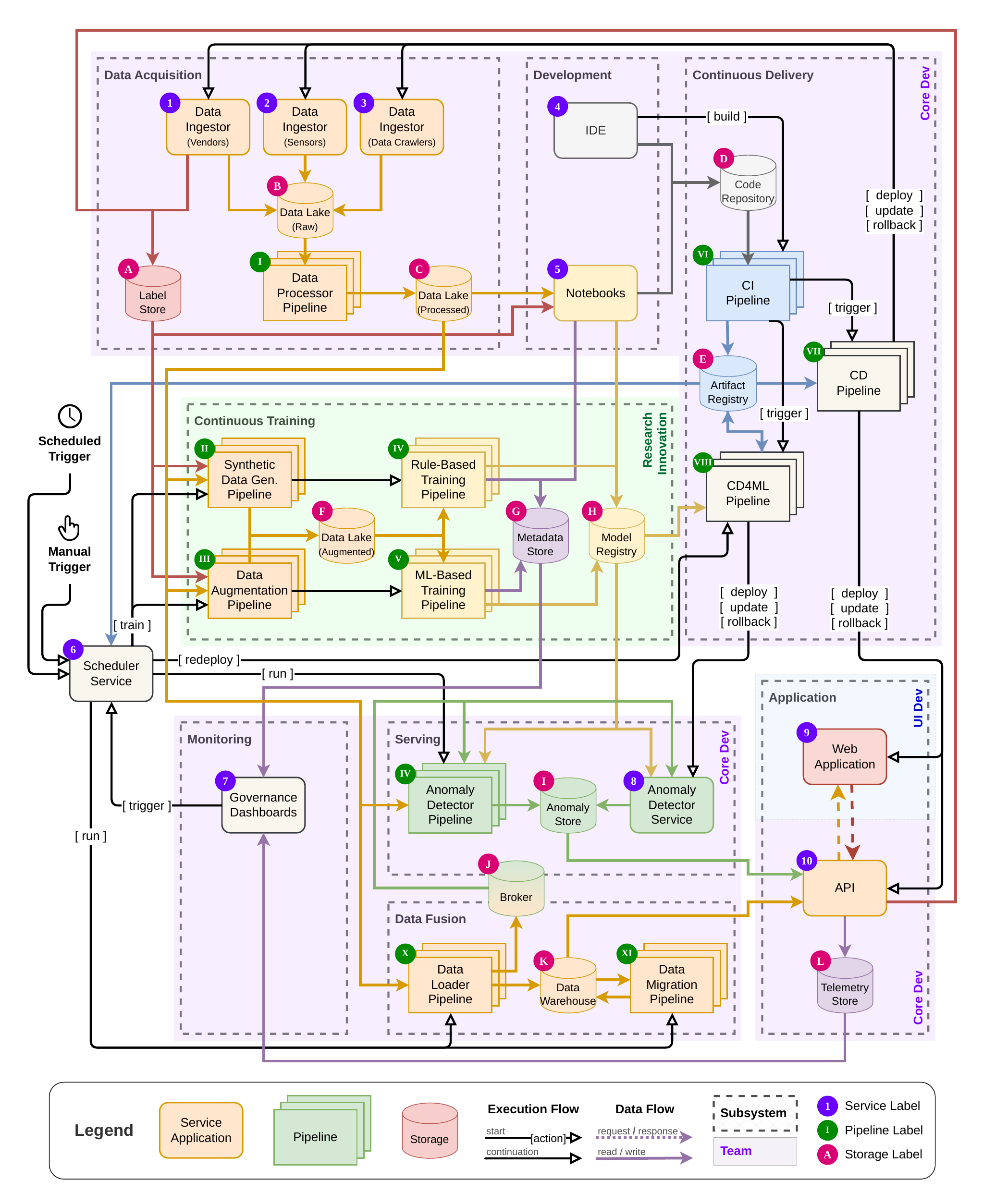}
  \caption{
    \textbf{System Architecture of the \OG tool.}
    The diagram follows the notation introduced by Ferreira%
    ~\cite{Ferreira:2025:PhD:MetricsOrientedArchitecturalModel},
    and is based on the previous publication about the \OG tool%
    ~\cite{Ferreira:2025:OceanGuard:MLOpsWithMicroservices}.
    Rectangles represent \textbf{services},
      which execute continuously.
    Stacked rectangles represent \textbf{pipelines},
      which execute on demand.
    Lastly, cylinders represent \textbf{data storage}.
    Components are connected by arrows.
    Black arrows with a hollow tip illustrate the \textbf{execution flow}.
      They start and end in a component.
      Labeled arrows represent the trigger that starts a workflow,
      whereas unlabeled arrows represent the continuation of an
      existing workflow.
    Colored arrows with a filled tip illustrate the \textbf{data flow}.
    They appear in two types:
      solid arrows going to and from a data storage represent
      write and read operations, respectively;
      dotted arrows represent a sync or async request-response
      communication between components.
    Components are colored according to the data they produce:
      \mbox{\LegendColoredComponent{orange}{data}},
      \mbox{\LegendColoredComponent{red}{labels}},
      \mbox{\LegendColoredComponent{gray}{source code}},
      \mbox{\LegendColoredComponent{blue}{executable artifacts}},
      \mbox{\LegendColoredComponent{yellow}{models}},
      \mbox{\LegendColoredComponent{green}{anomalies}}, and
      \mbox{\LegendColoredComponent{purple}{telemetry}}.
      The remaining \LegendBWComponent{standalone components} orchestrate
      the execution of others.
    Components are grouped into \textbf{subsystems},
    with backgrounds colored according to the \textbf{teams} responsible
    for their development.
    \LegendColoredLabel{violet}{Numbers},
    \LegendColoredLabel{moss}{roman numerals} and
    \LegendColoredLabel{magenta}{letters}
    are used as labels throughout \cref{sec:system_architecture}.
  }
  \label{fig:system_architecture}
\end{figure*}

The architecture of \OG is composed of multiple services required for the
functioning of the tool and follows the reference architecture proposed by
\citeauthor{Ferreira:2025:PhD:MetricsOrientedArchitecturalModel}%
~\cite{Ferreira:2025:PhD:MetricsOrientedArchitecturalModel}.
These are illustrated in \cref{fig:system_architecture}.

\OG microservices are grouped into seven subsystems, which interact via different
data and execution flows. The subsystems in the \OG tool are:
\begin{enumerate}
    \item \Subsystem{Data Acquisition},
          which collects data from multiple data sources and stores them for
          use by the system.
    \item \Subsystem{Continuous Training},
          which implements continuous training of the models
          used for anomaly detection.
    \item \Subsystem{Serving},
          which provides efficient anomaly detection
          using the aforementioned models.
    \item \Subsystem{Data Fusion}, 
          which organizes the data.
    \item \Subsystem{Application},
          which provides secure access to authorized users to operate and
          utilize the system.
    \item \Subsystem{Monitoring},
          which tracks the usage of the system.
    \item \Subsystem{Development},
          which facilitates the system development with pre-configured
          environments and tooling.
    \item \Subsystem{Continuous Delivery},
          which builds, tests, and deploys the components of the system.
\end{enumerate}

\section{System Architecture}
\label{sec:system_architecture}

\begin{figure*}[t]
  \centering
  \includegraphics[width=0.85\linewidth]{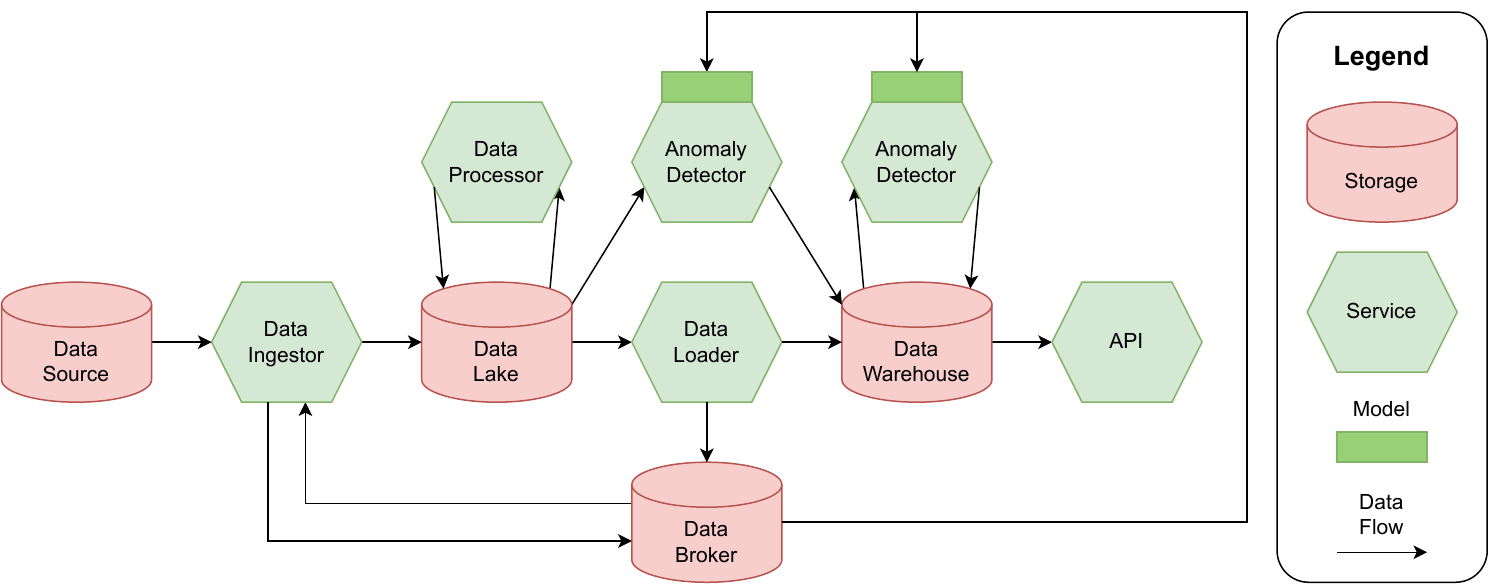}
  \caption{
    \textbf{Data flow of the \OG tool.}
    \LegendColoredLabel{red}{Red cylinders} represent data storage,
    while \LegendColoredLayer{green}{green hexagons} represent services.
    Arrows from a service to a storage represent writes,
    whereas arrows from a storage to a service represent reads.
  }
  \label{fig:data_flow}
\end{figure*}

This section presents the system architecture of the \OG tool. For convenience
of explanation, the section is divided according to the high-level data flow
shown in~\Cref{fig:data_flow}. However, the text refers to the components from
the software architecture diagram shown in~\Cref{fig:system_architecture}.

The reader is encouraged to explore \cref{fig:system_architecture} for a
detailed description of the \OG tool. For additional info about the requirements
that led to this architecture, please refer to our first paper about the
tool~\cite{Ferreira:2025:OceanGuard:MLOpsWithMicroservices}.
The implementation of these microservices will be discussed in detail in \Cref{sec:application_architecture}.

As described in \cref{sec:fundamentals}, the \OG tool is an MLES and
it follows the reference architecture proposed by%
~\citeauthor{Ferreira:2025:PhD:MetricsOrientedArchitecturalModel}%
~\cite{Ferreira:2025:PhD:MetricsOrientedArchitecturalModel}.
It consists of multiple microservices that help the tool achieve the goal of
anomaly detection. 

\subsection{Data Ingestor}\label{p:og_service_data_ingestor}
This service is responsible for data collection.
It is represented by one of the following components, which retrieve data from
different data sources:
  \LegendService{1}{3rd-Party Providers},
  \LegendService{2}{Physical Sensors}, and
  \LegendService{3}{Data Crawlers}.
Afterward, the service stores the data in the \LegendDataStore{B}{Data Lake (Raw)}, 
where it awaits further processing.

\subsection{Data Processor}\label{p:og_service_data_processor}
This service is responsible for data cleaning.
It is represented by the \LegendPipeline{I}{Data Processor Pipeline}, which
processes raw data from the \LegendDataStore{B}{Data Lake (Raw)} and stores
the results back into the \LegendDataStore{C}{Data Lake (Processed)}.

\subsection{Data Loader}\label{p:og_service_data_loader}
This service is responsible for data organization.
It is represented by the \LegendPipeline{X}{Data Loader Pipeline}, which 
retrieves processed data from the \LegendDataStore{C}{Data Lake (Processed)} and
loads it into the \LegendDataStore{K}{Data Warehouse}.
In addition, the service also requests anomaly detections by sending messages
via the \LegendDataStore{J}{Broker}.

\subsection{Anomaly Detector}\label{p:og_service_anomaly_detector}
This service is responsible for using ML-based anomaly detection.
It is represented by one of the following components, which implement different
patterns of model serving:
  \LegendPipeline{IV}{Anomaly Detector Pipeline}
  (for \emph{scheduled} prediction), and
  \LegendService{8}{Anomaly Detector Service}
  (for \emph{near real-time} prediction).
The service initializes its models from the \LegendDataStore{G}{Model Registry}
and stores the resulting anomalies into the \LegendDataStore{I}{Anomaly Store}.

\subsection{API}\label{p:og_service_api}
This service is responsible for exposing data for clients.
It is represented by the \LegendService{10}{API}, which
serves organized data from the \LegendDataStore{K}{Data Warehouse}
and anomalies from the \LegendDataStore{I}{Anomaly Store}.
In particular, the service can communicate with a \LegendService{9}{Web
Application} to provide a visual interface for clients.

\section{Application Architecture}
\label{sec:application_architecture}

This section dives deeper into the implementation of the microservices
introduced in \cref{sec:system_architecture}.

As described in \cref{sec:fundamentals}, the \OG tool is implemented using a
\Pattern{Hexagonal Architecture} pattern. The \Pattern{Core} of the \OG tool
defines the business logic based on the domain. It is independent of any
external dependencies, which are tied to the relevant \Pattern{Ports} that
provide abstract methods implemented through \Pattern{Adapters}.

\begin{figure*}[p]
  \centering
  \begin{subfigure}[t]{0.45\textwidth}
    \includegraphics[width=\linewidth]{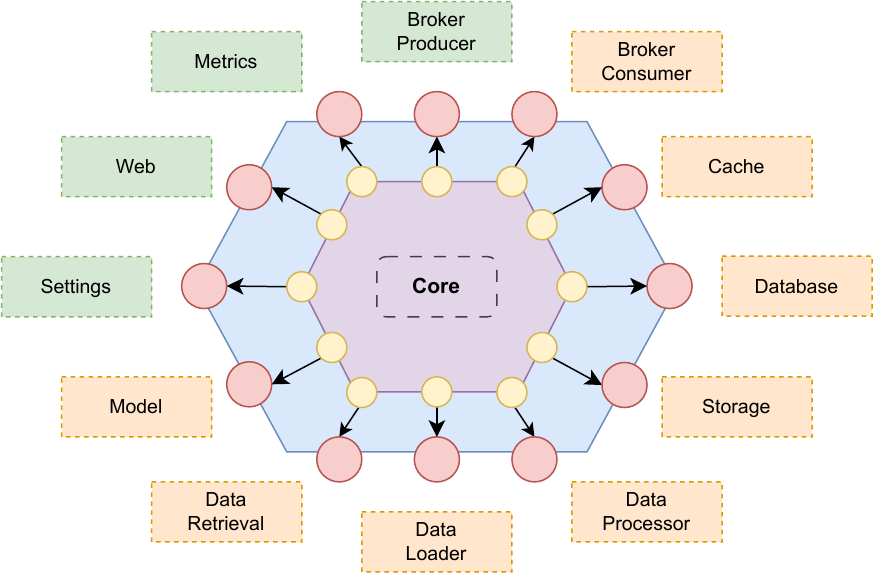}%
    \caption{
      Monorepo template
    }%
    \label{subfig:services_monorepo}
  \end{subfigure}%
  \hfill%
  \begin{subfigure}[t]{0.45\textwidth}
    \includegraphics[width=\linewidth]{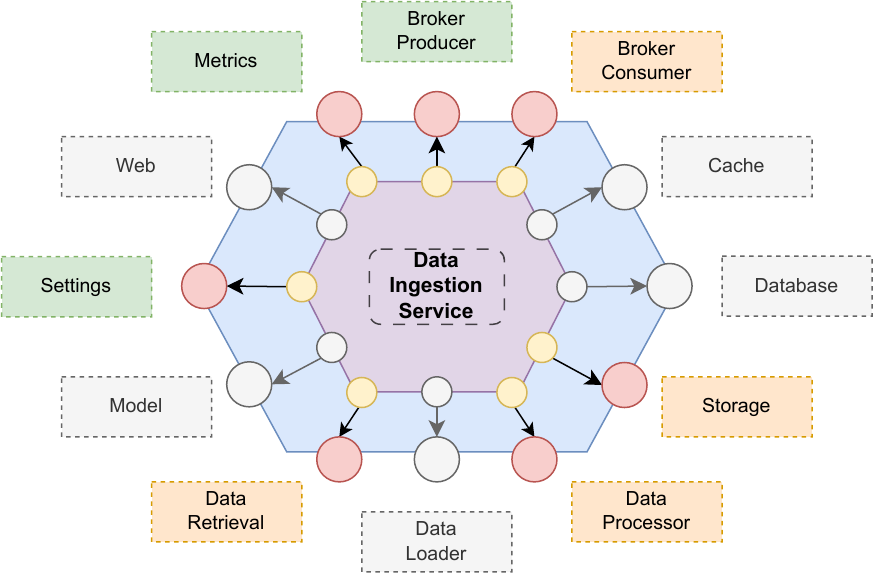}%
    \caption{
      \Service{Data Ingestor} service
    }%
    \label{subfig:services_data_ingestor}
  \end{subfigure}%
  \vspace{0.3cm}
  \begin{subfigure}[t]{0.45\textwidth}
    \includegraphics[width=\linewidth]{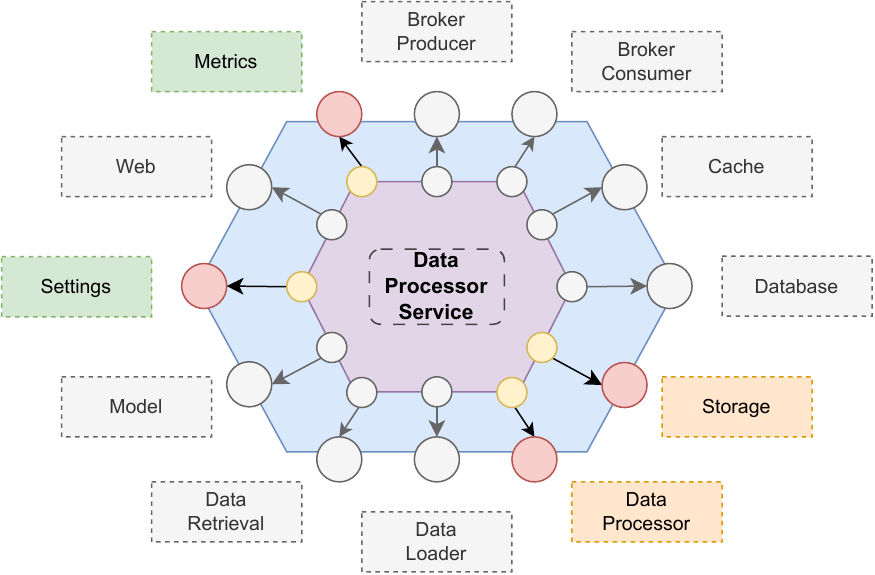}%
    \caption{
      \Service{Data Processor} service
    }%
    \label{subfig:services_data_processor}
  \end{subfigure}%
  \hfill%
  \begin{subfigure}[t]{0.45\textwidth}
    \includegraphics[width=\linewidth]{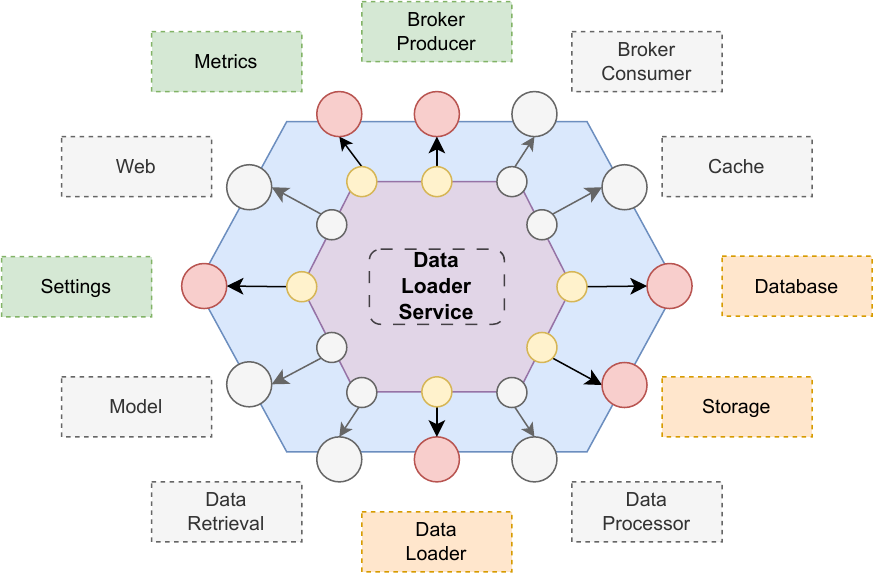}%
    \caption{
      \Service{Data Loader} service
    }%
    \label{subfig:services_data_loader}
  \end{subfigure}%
  \vspace{0.3cm}
  \begin{subfigure}[t]{0.45\textwidth}
    \includegraphics[width=\linewidth]{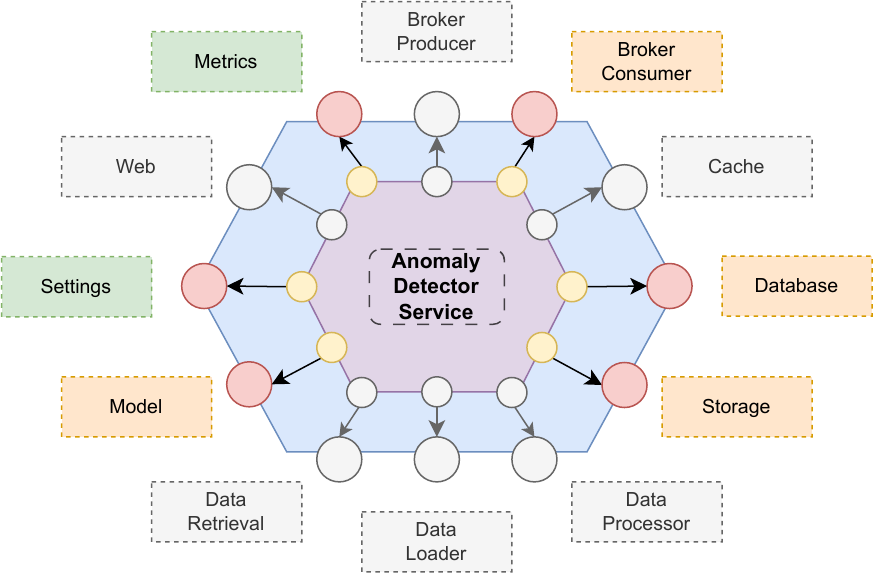}%
    \caption{
      \Service{Anomaly Detector} service
    }%
    \label{subfig:services_anomaly_detector}
  \end{subfigure}%
  \hfill%
  \begin{subfigure}[t]{0.45\textwidth}
    \includegraphics[width=\linewidth]{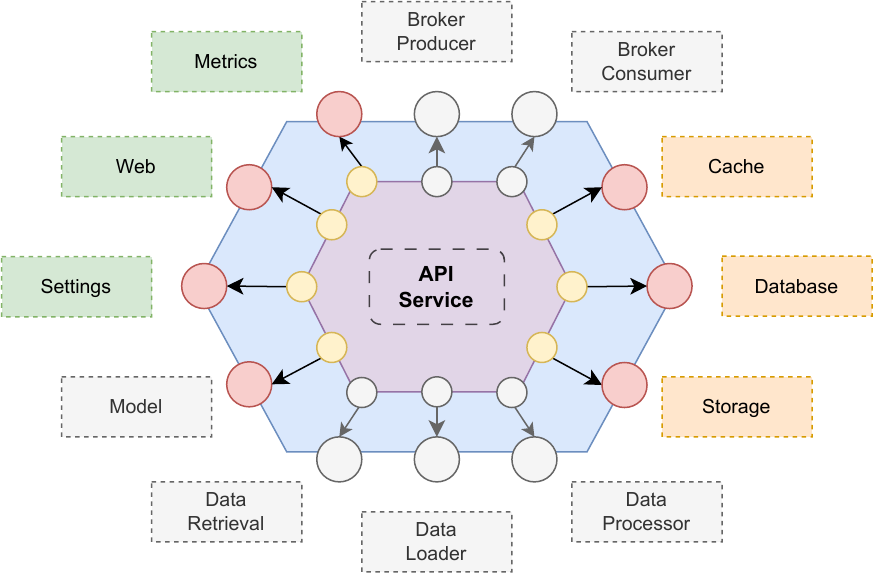}%
    \caption{
      \Service{API} service
    }%
    \label{subfig:services_api}
  \end{subfigure}%
  \vspace{0.3cm}
  \begin{subfigure}[t]{0.7\textwidth}
    \includegraphics[width=\linewidth]{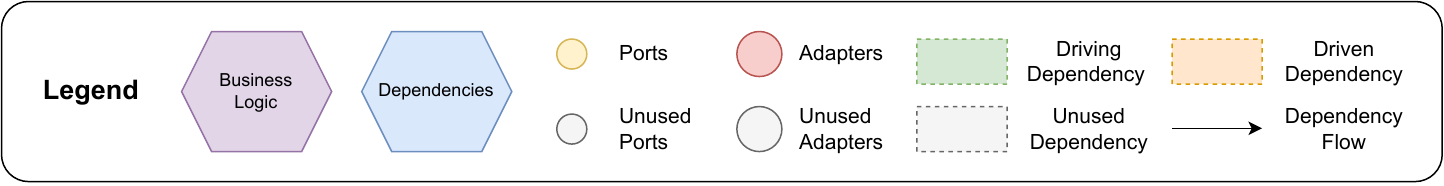}%
    \label{subfig:services_legend}
  \end{subfigure}
  
  \caption{%
    \textbf{Reusability in the \OG tool.}
    The monorepo template (\cref{subfig:services_monorepo})
    can be used to implement five types of microservices 
    (\crefrange{subfig:services_data_ingestor}{subfig:services_api})
    via the \Pattern{Ports and Adapters} pattern%
    ~\cite{Martin:CleanArchitecture:2017}.
    The \LegendColoredLayer{purple}{purple hexagon}
    represents the \textbf{Business Logic},
    where Domain-Driven Design patterns%
    ~\cite{Evans:DomainDrivenDesign:2003} are implemented.
    The \LegendColoredLayer{blue}{blue hexagon}
    represents the \textbf{Dependencies},
    where \LegendColoredLabel{yellow}{yellow circles} represent the \emph{Ports}
    and \LegendColoredLabel{red}{red circles} represent the \emph{Adapters}.
    Rectangles illustrate external dependencies,
    where a \LegendColoredComponent{green}{green rectangle}
    represents a \textbf{Driving Dependency}
    (which initiates actions in the service)
    while an \LegendColoredComponent{orange}{orange rectangle}
    represents a \textbf{Driven Dependency}
    (whose actions are initiated by the service). 
    Solid arrows, going from a port to an adapter represent the inside-out
    dependency flow, going from the \emph{Core} to the \emph{Ports} then
    to the \emph{Adapters}.
  }
  \label{fig:ports_adapters_reuse}
\end{figure*}

As illustrated by \cref{fig:data_flow}, the microservices introduced in
\cref{sec:system_architecture} have overlapping functionalities, in particular
in their communication with external storage components. To promote reusability
and avoid redundancy, the \OG tool is implemented in a single repository
(monorepo), allowing the sharing of common code between microservices.

\Cref{fig:ports_adapters_reuse} illustrates the reuse of \Pattern{Ports
and Adapters} through multiple microservices of the \OG tool.
The following external dependencies were supported:
\begin{itemize}
  \item \Adapter{Broker Producer}:
         Provides methods to produce messages in a broker queue.
  \item \Adapter{Broker Consumer}:
         Provides active and passive methods to listen and consume new messages
         from the broker queue.
  \item \Adapter{Cache}:
         Provides I/O methods to access cache.
  \item \Adapter{Database}:
         Provides various methods for performing I/O operations in a database.
  \item \Adapter{Data Loader}:
         Provides methods to load data into the data warehouse.
  \item \Adapter{Data Processor}:
         Provides methods to process raw data.
  \item \Adapter{Data Retrieval}:
         Provides methods to retrieve data from a given source.  
  \item \Adapter{Model}:
         Provides methods to perform predictions on a given model.
  \item \Adapter{Metrics}:
         Provides methods to monitor the performance of a service.
  \item \Adapter{Settings}:
         Provides methods to inject external dependencies into the services to
         enable execution.
  \item \Adapter{Storage}:
         Provides I/O methods to access a data lake.
  \item \Adapter{Web}:
         Provides methods to perform REST operations on the API.
\end{itemize}

Each microservice has its \textbf{core} business logic and uses a different
subset of available \Pattern{Ports and Adapters} to allow its implementation.
Moreover, each microservice has its main application to connect all
components and trigger their execution.

\medskip

The reusability through the use of a monorepo offers the possibility of
extending the \OG tool with new models, data types, external dependencies, and
ultimately new microservices. Furthermore, it allows the reuse of the tool for
future projects. Unfortunately, given the sensitivity of these projects, they
cannot be named.

An example source code demonstrating the implementation of the monorepo and
various microservices using the \Pattern{Ports and Adapters} is available in the
\nameref{sec:reproduction_package}.

\section{Challenges}
\label{sec:challenges}

This section describes two main challenges while applying the \Pattern{Ports and
Adapters} pattern and the monorepo-based solution during the development of the
\OG tool: \nameref{subsec:challenges:generality} and
\nameref{subsec:challenges:separation_of_concerns}.

\subsection{Generality}\label{subsec:challenges:generality}
Designing generic \Pattern{Ports and Adapters} that fit well in the context of
multiple microservices can be hard.
Ideally, a \Pattern{Port} should be specific, providing only what is necessary
to satisfy the functionalities required by the core business logic.
Moreover, it should be agnostic, ignoring details of any external dependencies
that might implement it.
These conditions are harder to satisfy when a \Pattern{Port} has multiple
\Pattern{Adapter} implementations.
Furthermore, the more services may reuse the same set of \Pattern{Port and
Adapters}, the harder it is to satisfy these conditions.

\subsection{Separation of Concerns}\label{subsec:challenges:separation_of_concerns}
Choosing when a snippet of code should be isolated into \Pattern{Ports and
Adapters} or become part of the \Pattern{Core} business logic can be hard.
Ideally, an \Pattern{Adapter} should be distinct, encapsulating a single
external dependency.
Moreover, it should be thin, providing no meaningful business logic.
business logic.
These conditions are harder to satisfy when an \Pattern{Adapter} handles
more complex, multi-step logic.
Furthermore, the more services have logic tightly coupled with external
dependencies, the harder it is to separate them.

\section{Lessons Learned}
\label{sec:lessons_learned}

This section describes three lessons learned during the development of the
\OG tool, related to the challenges described in
\Cref{sec:challenges}:
\nameref{subsec:lesson:compatibility},
\nameref{subsec:lesson:extensibility}, and
\nameref{subsec:lesson:incremental_development}.

\subsection{Compatibility}\label{subsec:lesson:compatibility}
As described in \cref{sec:system_architecture}, the \OG services have
overlapping functionalities. By reusing the same \Pattern{Ports and Adapters}
between microservices that communicate with the same storage, it is easier
to make them compatible with the external dependencies.

\subsection{Extensibility}\label{subsec:lesson:extensibility}
As described in \cref{sec:system_architecture}, the \OG services fulfill
different roles in the system. By reusing the \Pattern{Ports and Adapters}
and isolating each individual \Pattern{Core}, it is easier to create
new microservices that provide different functionalities.

\subsection{Incremental Development}\label{subsec:lesson:incremental_development}
As described in \cref{sec:system_architecture}, the \OG services compose a
complex MLES. By adopting the \Pattern{Ports and Adapters} pattern to implement
its microservices, it is easier to incrementally develop and update the 
\OG tool, both on the architecture and application level.

\section{Conclusion}
\label{sec:conclusion}

The \OG tool is an extensible MLES whose aim is to analyze and detect anomalies
across multiple types of data from the maritime domain
~\cite{Ferreira:2025:OceanGuard:MLOpsWithMicroservices}.
This experience report described the implementation of various microservices in
the \OG tool. In particular, it highlighted the benefits of reusing the
\Pattern{Ports and Adapters} pattern to build them from a single codebase.

During development, the authors faced two major challenges to ensure
reusability:
\begin{itemize}
  \item \emph{generality},
         related to defining \Pattern{Ports} that are specific
         and dependency-agnostic, and
  \item \emph{separation of concerns},
         related to defining \Pattern{Adapters} that are distinct
         and logic-thin.
\end{itemize}

Furthermore, the authors learned three lessons related to the benefits of
reusing the \Pattern{Ports and Adapters} pattern while sharing code via a
monorepo:
\\
\begin{itemize}
  \item \emph{compatibility} between microservices that communicate with
         the same external dependencies,
  \item \emph{extensibility} to new microservices that may reuse the same
         external dependencies, and
  \item \emph{incremental development} of microservices as new external
         dependencies get supported.
\end{itemize}

Future work on the development of the \OG tool will focus on developing and
extending all the components required by (re)using the implementation approach
described in this paper.
The development process will be guided by the fundamentals defined in
\Cref{sec:fundamentals}.

\section*{Reproduction Package}
\label{sec:reproduction_package}

An example source code that illustrates the software architecture practices
described in this paper is available at
\href{https://www.github.com/ocean-guard/pirate-guard}{github.com/ocean-guard/pirate-guard}.
The code contains a simplified implementation of the \Pattern{Ports and
Adapters} introduced in \cref{sec:application_architecture}, as well as
an example of \Pattern{Core} business logic.
The implementation satisfies the five core properties of a robust scientific
contribution: it is re-runnable, repeatable, reproducible, reusable, and
replicable~\cite{Benureau:2018:TransformingCodeScience}.


\printbibliography

\end{document}